\documentclass[twocolumn,aps,pra,showpacs,superscriptaddress,
longbibliography,groupedaddress,10pt]{revtex4-1}

\usepackage{bm}
\usepackage{amssymb}
\usepackage[usenames,dvipsnames]{color}
\usepackage{graphicx}
\usepackage{dcolumn}
\usepackage{hyperref}
\usepackage{natbib}
\usepackage[caption=false]{subfig}
\usepackage{float}
\usepackage{times}
\usepackage{booktabs}		
\usepackage{mathtools}
\usepackage{amsmath,array}
\usepackage{relsize}
\usepackage{enumerate}
\usepackage{color}
\usepackage{pdfpages}

\usepackage{listings}
\usepackage{epstopdf}
\usepackage{braket}

\begin{document}

\title{Same Period Emission and Recombination in Non-Sequential Double Recombination High-Harmonic Generation in H$_2$}
\author{Kenneth K. Hansen}
\author{Lars Bojer Madsen}
\affiliation{Department of Physics and Astronomy, Aarhus University, DK-8000, Denmark}

\pacs{42.65.Ky,33.80.Rv,34.80.Bm}

\begin{abstract}
Non-sequential double recombination (NSDR) high-order harmonic generation (HHG) is studied in a molecular system.
We observe a unique molecular two-electron effect with a characteristic cutoff in the HHG 
spectrum at higher energies than what was previously  seen for NSDR HHG in atoms.
The effect is corroborated with a classical model where it is found that the effect is sensitive to the molecular potential 
and originates from same period emission and recombination (SPEaR) of two electrons.
The effect persists for intermediate nuclear distances of $R \gtrsim 8.0$ a.u.
\end{abstract}
\date{\today}
\maketitle

\section{Introduction}\label{intro}
Through the emergence of intense laser pulses, high-harmonic generation (HHG) has arisen as both a method of creating ultra-short 
pulses and as a tool to probe electron dynamics in atomic and molecular systems. Using a many-cycle pulse of linearly polarized light HHG 
cutoff energies of $I_p + 3.17U_p$ have been found for one-electron dynamics, where $I_p$ is the ionization potential and 
$U_p=I/(4 \omega^2)$ is the ponderomotive potential with $I$ the laser intensity and $\omega$ the angular 
frequency~\cite{Krause_1992,Corkum_HHG,Lewenstein_HHG} (Atomic units are used throughout unless indicated otherwise).
Molecular HHG spectra show a characteristic two-center interference minimum \cite{mol_interferens}, and control of alignment of the molecular axis with respect to the polarization of the driving field leads to harmonics of non-linear ellipticity \cite{alignment_theory,alignment_experiment}. In relation to this work, which focuses on a unique molecular two-electron mechanism with a distinctive cutoff, we mention that for very large internuclear distances cutoffs in the HHG spectra emerge from the propagation of a single electron directly from one nuclei to another leading to kinetic energies  up to $8U_p$\cite{Bandrauk,MLein}.

Recently a two-electron HHG process was reported for an atomic system \cite{NSDR}. This non-sequential double recombination (NSDR) HHG process results in a new plateau in the HHG spectrum reaching beyond the one-electron HHG signal.
NSDR can be explained as two electrons propagating independently of each other in the field and then returning at the same instant to emit the combined kinetic energy of the two electrons as HHG. It was found that because electrons emitted in the same half-period would have to propagate the same path to return at the same instant, electron-electron repulsion would make such a same-period emission process highly unlikely in the atomic case. Instead, only electrons emitted in different half-periods of the pulse could return at the same instant without electron-electron repulsion  suppressing the process. Such 
electrons emitted in different periods traverse the nuclei more than once and can reach combined maximum return kinetic energies of $5.55U_p = 3.17U_p + 2.38U_p$ and $4.70U_p = 3.17U_p + 1.53U_p$ for first and third electronic return combined and first and second electronic return combined, respectively.
To our knowledge the manifestation of NSDR HHG  in a molecular system has not been studied, and it is the purpose of the present work to do so.

In the molecular case, we find that for internuclear distances of $R\gtrsim 8.0$ a characteristic signal is observed with even higher energy than that of atomic NSDR HHG. This signal and the associated cutoff is identified to stem from electrons which both are emitted and recombine within the same period. This same period emission and recombination (SPEaR) allows for a higher total kinetic energy of the electrons than what is allowed for an atomic system where electron-electron repulsion suppresses such a signal. This conclusion is supported by a classical analysis, a short-time Fourier transform analysis and the behavior of the signal in the long pulse limit.

The paper is organized as follows. In Sec. \ref{sec:methods}, we describe the numerical methods used. In Sec. \ref{sec:results}, we present results and in Sec. \ref{sec:conclusion}, we conclude.

\section{Numerical Methods}\label{sec:methods}
Time-dependent Schr\"{o}dinger equation (TDSE) calculations are made using a co-linear model for H$_2$. 
The Hamiltonian reads
\begin{align}
H(t) = \sum^2_{i=1} \left(\frac{[p_i + A(t)]^2}{2} + V_R(x_i) \right) + W(x_1-x_2),
\end{align}
with $p_i$  the canonical momentum, $A(t)$ the vector potential, $V_R(x_i) = -Z ((R/2-x_i)^2 + \epsilon_{ei})^{-1/2}-Z ((R/2+x_i)^2 + \epsilon_{ei})^{-1/2}$ the Coulomb interaction with the nuclei, $Z$ the nuclear charge, $R$  the intermolecular distance, $x_i$  the position of electron $i$ and $W(x_1-x_2) = ((x_1-x_2)^2 + \epsilon_{ee})^{-1/2}$ the electron-electron interaction. Our model He system is obtained by setting $R=0$ in the above. 
The softcore parameters were set to $\epsilon_{ei} = 0.5$ and $\epsilon_{ee} = 0.329$ corresponding to $I_p^{(1)} = 0.9$ and $I_p^{(2)} = 2.0$ for ionization of respectively the first and second electron in He. 
We keep the softening parameters fixed in all calculations. For $R=16$, e.g., 
the ionization potential is $I_p^{(1)} = 0.95$ and $I_p^{(2)} = 0.89$.
The TDSE
is solved using a split-step operator Crank-Nicolson (Peaceman-Rachford) method as in Ref. \citep{NSDR}.
The harmonic spectrum is calculated by taking the modulus square of the Fourier transformed dipole acceleration, $a_{dip}(t)$, which is calculated in every time-step~\cite{dipole_accel}. Ground states and their energies are found by imaginary time-propagation.
A grid step-size of $\Delta x = 0.2$ is used in a box of size $L = 400$ symmetric around 0, an imaginary time-propagation step-size of $\Delta t = 0.15$ with a 1000 steps and real time propagation step-size of $\Delta t = 0.075$ ensured convergence.
The pulse used has the form $A(t) = \frac{F_0}{\omega} \sin^2 \left(\frac{\omega t}{2n}\right) \sin (\omega t), \quad 0\leq t \leq T_n$ with typically 6-cycles ($n=6$), $\omega = 0.0584$ ($\lambda \simeq 780$ nm), $F_0 = 0.119$ $(I\simeq 5.0\times 10^{14}$ W/cm$^{2})$ and the pulse polarization is linear and parallel with the molecular axis.
Using a short pulse modifies the maximum kinetic return energy an electron can obtain from the electric field. Classical cutoff energies for first, second and third return of an electron are given in Table \ref{tab:cutoff} in the long pulse limit and for  the considered 6-cycle pulse. The two-electron entries 
with first and second returns and with first and third returns give the cutoffs reported for  NSDR in the atomic case previously~\cite{NSDR}.  As discussed in Sec.~I, 
the possibility for the  two-electron event where both electrons recombine at the first return (last row in Table I) is suppressed by electron-electron interaction in the atomic case.  This two-electron process is not suppressed from moderately large internuclear distances in the molecular case and 
is the origin of SPEaR NSDR HHG as explained in 
the next section.

 \begin{table}
 \caption{\label{tab:cutoff}Maximum electron return kinetic energies in units of $U_p$ in the long pulse limit and for the 6-cycle pulse for one 
 and two electrons. The left column denotes the number of electrons involved and the return events.
 }
  \begin{ruledtabular}
 \begin{tabular}{l r r}
	&	long pulse &	6-cycle pulse \\
\hline
one-electron, first return	&	$3.17$	&	$3.08$\\
one-electron, second return &	$1.53$	&	$1.61$\\
one-electron, third return &	$2.38$	&	$2.15$\\
two-electrons, first and second	return &	$4.70$	&	$4.69$\\
two-electrons, first and third	return &	$5.55$	&	$5.23$ \\
two-electrons, first and first	return&	$6.34$	&	$6.16$
\end{tabular}
 \end{ruledtabular}
 \end{table}

\begin{figure}
\centering
\includegraphics[width=\columnwidth]{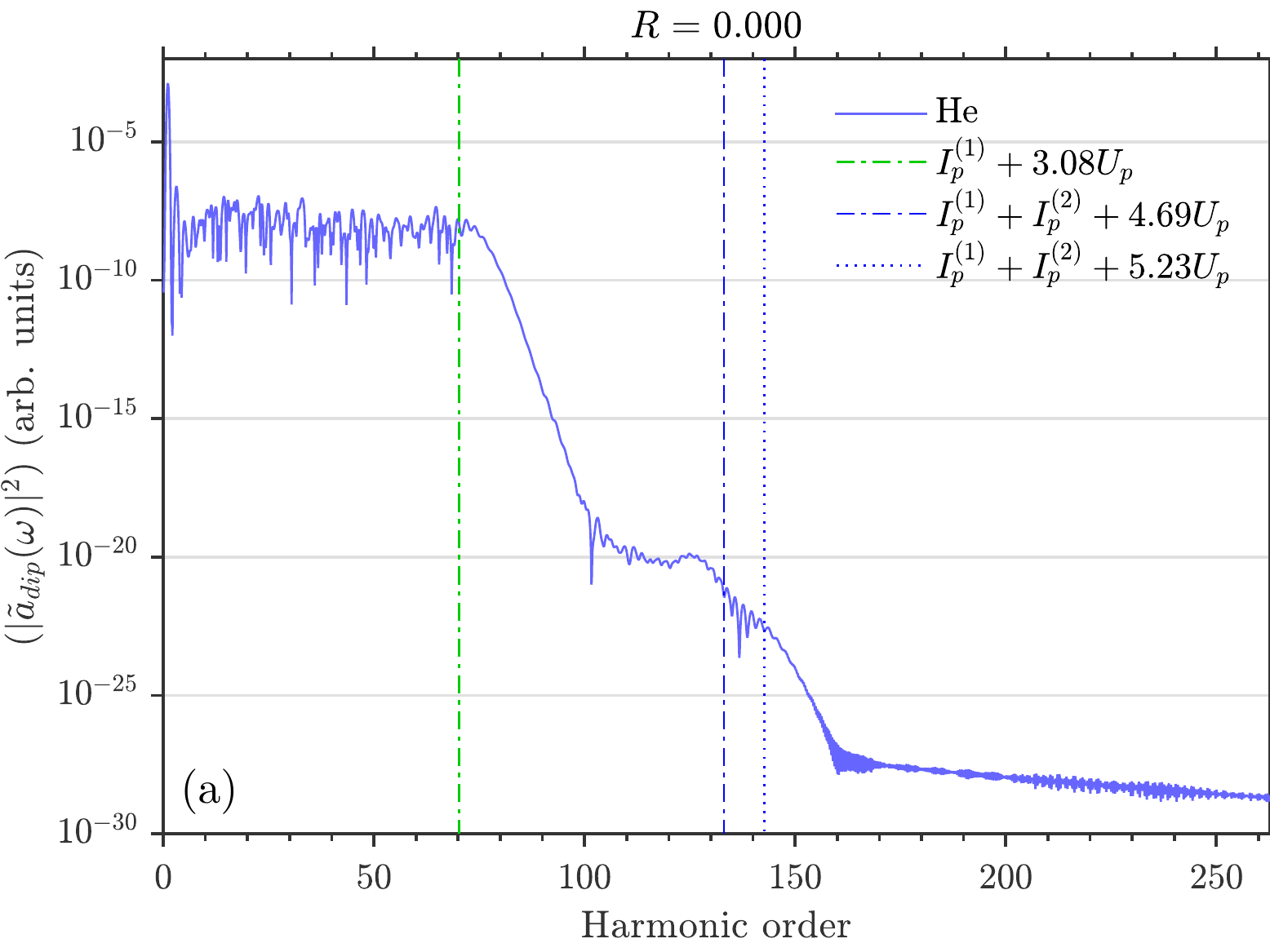}
\includegraphics[width=\columnwidth]{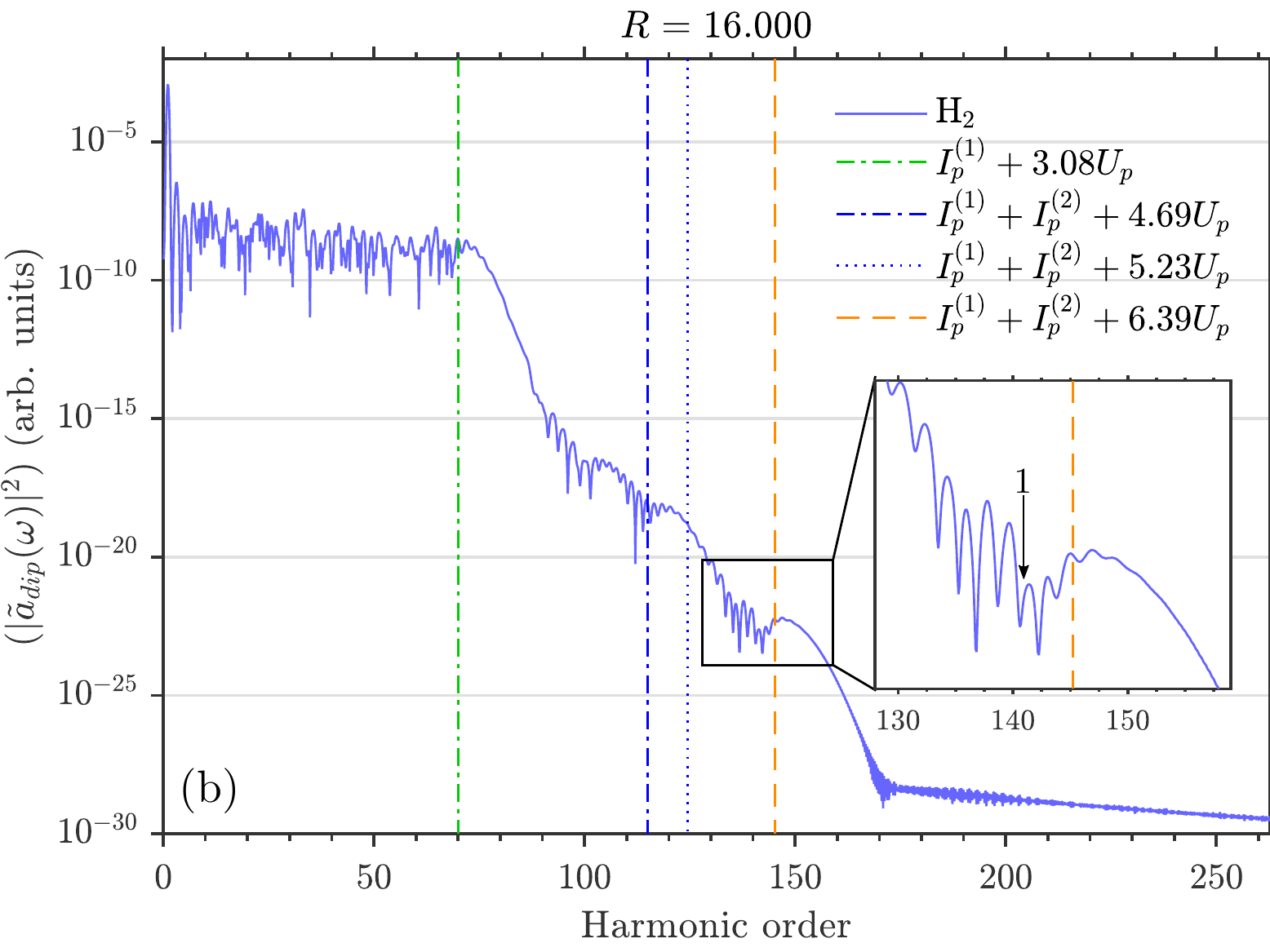}
\caption{(Color online) HHG spectra for (a) He and (b) H$_2$ with $R=16.0$.
The spectra are shown in units of the Harmonic order for an $n=6$-cycle laser pulse at $\omega = 0.0584$ ($\lambda \simeq 780$ nm) and $F_0 = 0.119$ $(I\simeq 5.0\times 10^{14}$ W/ cm$^{2})$. The HHG cutoffs for one- and two electron HHG are shown as the dotted vertical lines at $I_p^{(1)} + 3.08U_p$, $I_p^{(1)} + I_{p}^{(2)} + 4.69U_p$ and $I_p^{(1)} + I_{p}^{(2)} + 5.23U_p$. In (b) the SPEaR NSDR HHG cutoff is shown as the dashed line at $I_p^{(1)} + I_{p}^{(2)} + 6.39U_p$. The insert in (b) shows a zoom-in of the SPEaR NSDR cutoff. The arrow at 1 is at $I_p^{(1)} + I_{p}^{(2)} + 6.16U_p$ (see text).}
\label{fig:spectre}
\end{figure}

\section{Results and Discussions}\label{sec:results}
Our co-linear model allowed us to perform calculations for many different $R$'s in the interval $[0; 80]$.
Figure \ref{fig:spectre} shows the HHG spectra for our model He [Fig.~\ref{fig:spectre}(a)] and H$_2$ [Fig.~\ref{fig:spectre}(b)] with internuclear distance $R=16.0$ as a representative example.
For the 6-cycle pulse, 
Fig.~\ref{fig:spectre}(a) shows the one-electron cutoff of $I_p^{(1)} +3.08U_p$ and the atomic NSDR cutoffs $I_p^{(1)} + I_p^{(2)} +4.69U_p$ and $I_p^{(1)} + I_p^{(2)} +5.23U_p$ [see Table I].
Comparing the cutoffs in Figs.~\ref{fig:spectre}(a) and (b) a new cutoff can be seen in Fig.~\ref{fig:spectre}(b) at $\Omega \approx 145 \omega$. 
Our extensive calculations for many $R$'s show that this new cutoff emerges for internuclear distances of $R\gtrsim  8.0$, i.e., in the limit
where the Born-Oppenheimer potential is relatively flat. In this limit, the two electrons are predominantly confined to separate nuclei and the electron-electron interaction is weak.
When we increase $R$ from $R=0$ (He case) up to $R \simeq 8$, we observe a continuous change of the atomic-like NSDR signal. At $R=8$, we see 
a pronounced build-up of the new  signal seen in Fig.~\ref{fig:spectre}(b).

For the present laser parameters, the one-electron signal with a cutoff of up-to $8 U_p$ and corresponding to direct trajectories from one nuclei to another
becomes dominant for $R \gtrsim 40.0$, and therefore the new signal is only observed for $R \in[8; 40]$. 
Comparing this new cutoff with the atomic NSDR cutoffs, no similarity can be observed for all internuclear distances where the new cutoff is observed. Also the new cutoff does not have the distinct strong dependence on the internuclear distance expected for the one-electron high energy cutoff for large internuclear distances \cite{Bandrauk,MLein}. We therefore conclude that it can not originate from earlier proposed one-electron or two-electron HHG mechanisms. 
Here we propose a unique molecular two-electron process of NSDR HHG to explain this new cutoff.

\begin{figure}
\centering
\includegraphics[width=\columnwidth]{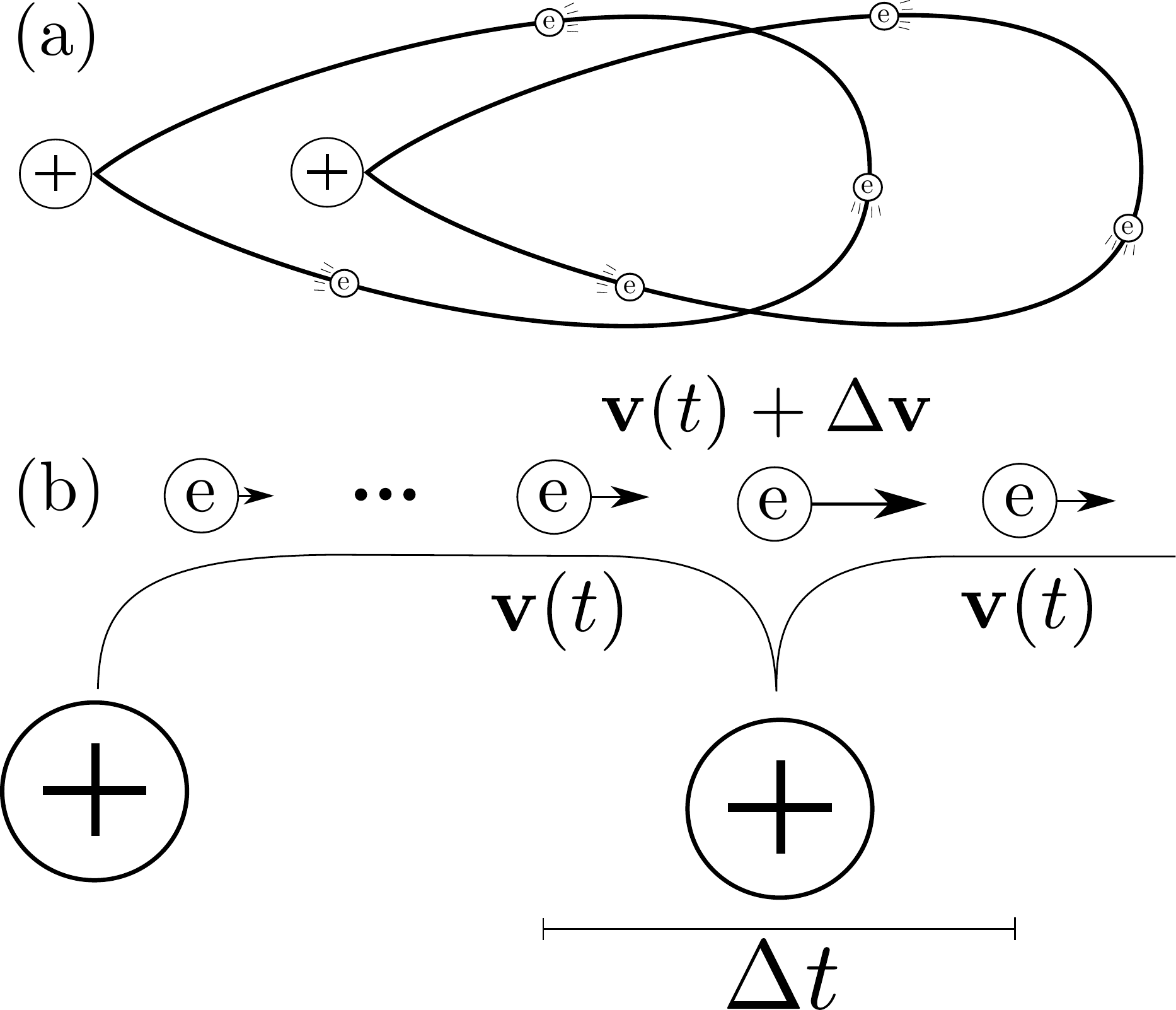}
\caption{(a) Illustration of the mechanism responsible for SPEaR. At large intermolecular distances, $R\gtrsim 8$, electrons emitted at the same time at different nuclei will interact relatively weakly. Therefore both electrons are able to reach the maximum kinetic return energy of $3.08U_p$ in the considered pulse. (b) An electron emitted at one nuclei in NSDR (both electrons are in the continuum for NSDR) could traverse a bare nuclei and thereby obtain a velocity increase of $\Delta \mathbf{v}$ which leads to a longer path in the continuum of approximately $\Delta \mathbf{r} = \Delta \mathbf{v} \Delta t$. This increases the kinetic energy of the returning electron by an amount dependent on the intermolecular distance (See text).}
\label{fig:SPEaR_classical}
\end{figure}

\begin{figure}
\centering
\includegraphics[width=\columnwidth]{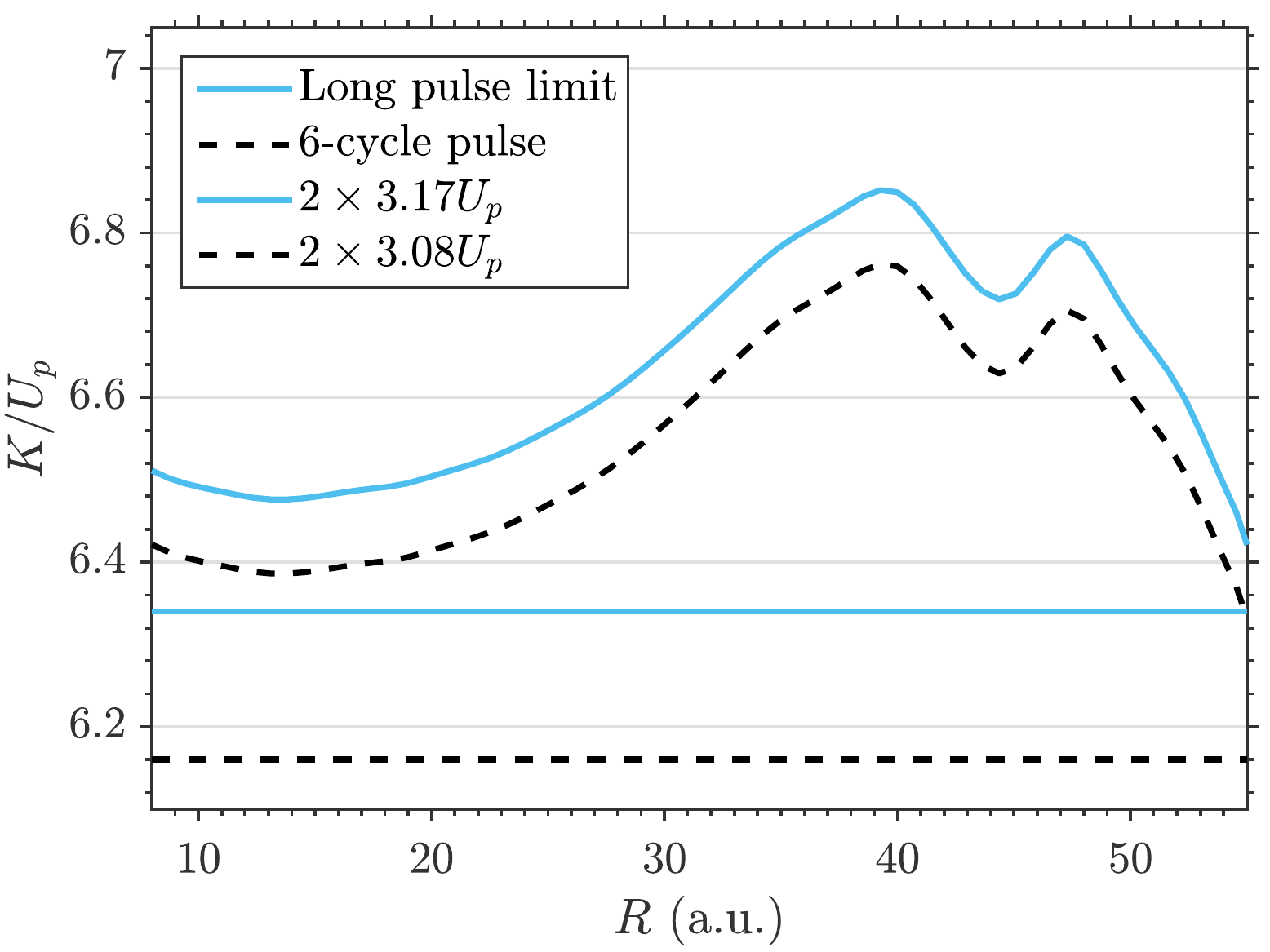}
\caption{(Color online) Classical SPEaR return kinetic energy $K$ in units of $U_p$ of one electron propagating in a laser field added to an electron propagating in a laser field and a Coulomb potential located at $R$ with emission at $r(t_i)=0$ as a function of the intermolecular distance, $R$. Here the combined kinetic return energy of the two electrons is shown for the long pulse limit (upper solid curve) and for the 6-cycle pulse (upper dashed curve) and the expected SPEaR return kinetic energy without the modification introduced by the Coulomb potential for the long pulse limit (lower solid line) and the 6-cycle pulse (lower dashed line).}
\label{fig:SPEaR_cutoff}
\end{figure}

As the internuclear distance increases the assumption that electron-electron repulsion forbids electrons emitted in the same period from returning at the same instant and recombining as NSDR is no longer justified.
For large $R$, the electron-electron repulsion plays a minor role even if the electrons are emitted at the same instant.  Therefore they will be able to propagate approximately the same path as if they where emitted from an atomic system and return with the combined maximum return kinetic energy of $2\times3.08U_p = 6.16U_p$ for the 6-cycle pulse ($2\times3.17U_p = 6.34U_p$ in the long pulse limit) [see Table I].
This is same period emission and recombination NSDR (SPEaR NSDR) and the mechanism is illustrated in Fig. \ref{fig:SPEaR_classical}(a).

Because the molecule is oriented with the molecular axis  parallel to the laser field polarization, the cutoff for SPEaR  depends on the internuclear distance.
One electron interacts with a bare nuclei as shown in Fig. \ref{fig:SPEaR_classical}(b) for high return kinetic energy paths. This interaction modifies the electron propagation for one of the electrons, and enables an increase of the kinetic energy of the returning electron and makes the cutoff dependent on the internuclear distance.

The SPEaR NSDR cutoff as a function of the internuclear distance is approximated by modeling a single electron propagating in the pulsed-field interacting with the Coulomb potential. The cutoff energy is then calculated by solving the classical equations of motion numerically and adding the atomic HHG cutoff energy to the maximum return kinetic energy of the modeled electron.
The classical cutoff predicted by this model is found to fit for all 
calculations made in the range $R\in [8.0;40.0]$  where the new signal is observed. A mentioned above, for $R \gtrsim 40.0$ the SPEaR NSDR HHG signal is not observed as the one-electron cutoff from direct paths becomes the dominant signal in that spectral range by several orders of magnitude. 
The difference between the simple model of the cutoff, $I_p^{(1)} + I_p^{(2)} + 6.16U_p$, [marked at 1 in Fig.~\ref{fig:spectre} (b)] and the model with the Coulomb potential included, $I_p^{(1)} + I_p^{(2)} + 6.39U_p$, is highlighted in the insert in Fig. \ref{fig:spectre} (b). Though the difference is relatively small the inclusion of the Coulomb potential results in the classical prediction of the cutoff fitting perfectly for all internuclear distances that we have considered. This is in contrast to the simple model where the predicted cutoff always is shifted away from the observed cutoff.
The cutoff as a function of the internuclear distance is shown in Fig. \ref{fig:SPEaR_cutoff}.
The calculated cutoffs shown in Fig. \ref{fig:SPEaR_cutoff} are dependent on the pulse parameters as the specific excursion path and excursion length of the electron is dependent on the pulse parameters and the interaction is an interplay between the specific excursion path and excursion length of the electron and the internuclear distance.
The structure in the calculated SPEaR NSDR cutoffs in Fig.~\ref{fig:SPEaR_cutoff} around $R \approx [40;50]$ is due to this being the maximum propagation length for short HHG paths.

We also performed calculations in the long pulse limit and found that SPEaR NSDR HHG is still observed and that the classically predicted cutoff is correct.

\begin{figure}
\centering
\includegraphics[width=\columnwidth]{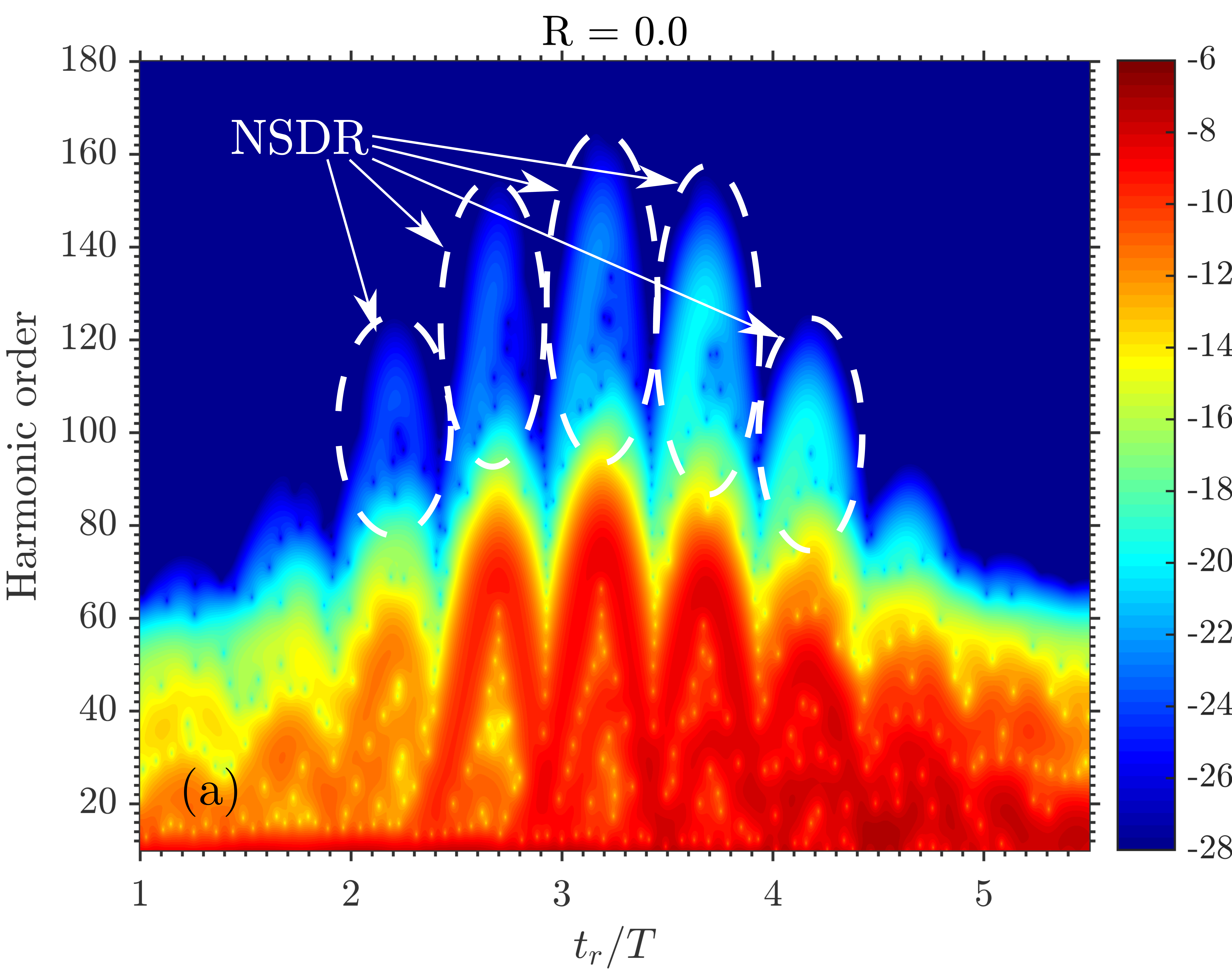}
\includegraphics[width=\columnwidth]{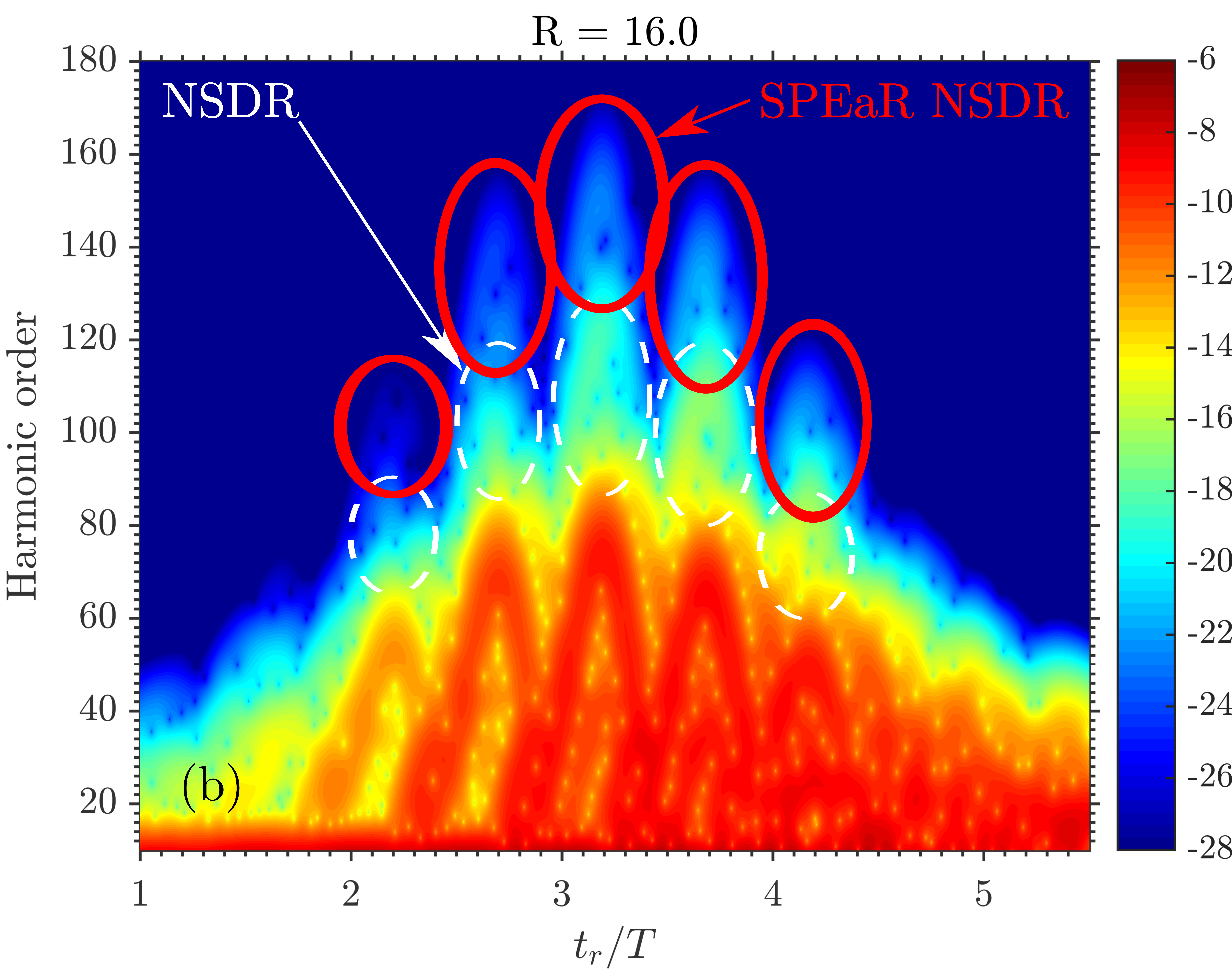}
\caption{(Color online) Norm square of the short-time Fourier transforms of the dipole acceleration of our model He and H$_2$ with R=16.0 on a logarithmic scale as a function of the return time $t_r$ in units of the laser cycle period, $T$, and return energy in orders of the pulse center frequency $\omega$. The NSDR signal is marked with dashed circles in (a) and (b). SPEaR NSDR HHG is marked with full circles in (b).}
\label{fig:gabor}
\end{figure}

To compare the classical model with quantum mechanical results a short-time Fourier transform (STFT) (Gabor transform) of the dipole acceleration is performed:
\begin{align}
\tilde{a}_{dip} (\Omega,t) = \int_{-\infty}^{\infty} \mathrm{d} t' \, e^{-i\Omega t'} a_{dip}(t) e^{-4 \ln (2) (t'-t)^2/\mathrm{FWHM}^2},
\end{align}
where FWHM is the full-width-half-maximum duration of the Gaussian window function which is set to $0.1\times2\pi/\Omega$.
Figures \ref{fig:gabor}(a) and (b) show the STFT of the dipole acceleration for our model He and H$_2$ for $R=16.0$. The NSDR signal is marked with dashed circles in both figures and the SPEaR NSDR signal is marked with full circles in Fig.~\ref{fig:gabor}(b).
The SPEaR NSDR is clearly located directly above the NSDR signal and not shifted to the left as would be expected for the high energy cutoffs observed for large internuclear distances. This shift to the left is expected because the high energy cutoff for large internuclear distances originates from direct exchange paths where electrons propagate directly from one nuclei to the other. Therefore the recombination time will happen earlier than for electrons propagating out and returning again for recombination.
The new signal is also seen directly above one-electron HHG which supports the conclusion that this new signal originates from a similar process to one-electron HHG and NSDR HHG where the electron propagates in the field and returns to the same nuclei after changing direction in the continuum, as described by the three-step-model.
The SPEaR signal being above the one-electron HHG and NSDR HHG is also what is expected form the classical model and therefore 
the STFT results support the proposed model.

\section{Summary and Outlook}\label{sec:conclusion}
In this work we have investigated NSDR HHG for a homo-nuclear model of H$_2$.
We have performed calculations over an extended range of internuclear distances and have found that a new  effect arises for $R \gtrsim 8.0$ and persists until the high energy one-electron cutoff from electrons propagating directly from one nuclei to another reaching of up-to $8U_p$ dominates the frequency range.
A classical model is proposed to explain this new cutoff. Electrons emitted and recombining in the same period (SPEaR NSDR HHG) are found to predict the observed cutoffs and the results are found to extend to the long pulse limit. It is also found that an interaction with the bare nuclei has to be included to correctly predict the cutoffs observed in the TDSE HHG spectra.
This interaction should be included for precise predictions of cutoffs for all processes where electrons traverse a bare nuclei and further work on molecular NSDR should therefore include this interaction when calculating the return kinetic energies of electron returning to a nucleus for the second or third time.
We also performed a short-time Fourier transform and found that SPEaR NSDR HHG is emitted similar to one-electron HHG and NSDR HHG. This is what is predicted by the classsical model for SPEaR NSDR and therefore additionally supports the mechanism proposed in this paper.

\section*{Acknowlegdements}
This work was supported by the ERC-StG (Project No. 277767-TDMET) and the VKR Center of Excellence.
We thank Dr. D. Bauer for making the two-electron TDSE program with which much of the calculations were performed available.
The numerical results presented in this work were performed at the Centre for Scientific Computing, Aarhus \url{http://phys.au.dk/forskning/cscaa/}.

%

\end{document}